\documentclass[conference]{IEEEtran}
\IEEEoverridecommandlockouts
\usepackage{cite}
\usepackage{amsmath,amssymb,amsfonts}
\usepackage{algorithmic}
\usepackage{graphicx}
\usepackage{textcomp}
\usepackage{xcolor}

\usepackage[outdir=./]{epstopdf}

\usepackage{enumitem}
\newlist{steps}{enumerate}{1}
\setlist[steps, 1]{label = Step \arabic*:}
\allowdisplaybreaks
\def\BibTeX{{\rm B\kern-.05em{\sc i\kern-.025em b}\kern-.08em
\usepackage{hyperref}
    T\kern-.1667em\lower.7ex\hbox{E}\kern-.125emX}}
    
\DeclareMathOperator*{\argmax}{argmax}

\begin{document}

\def\mycmd{1} 
\def\mycmdPeriodogram{1} 

\title{A Framework for UAV-based Distributed Sensing Under Half-Duplex Operation 
\author{\IEEEauthorblockN{Xavier A. Flores Cabezas; Diana P. Moya Osorio; and Markku Juntti}
\IEEEauthorblockA{Centre for Wireless Communications, University of Oulu, Finland}
\IEEEauthorblockA{Email: [xavier.florescabezas, diana.moyaosorio, markku.juntti]@oulu.fi}\vspace{-2em}}
}
%




\maketitle

\begin{abstract}

This paper proposes an unmanned aerial vehicle (UAV)-based distributed sensing framework that uses frequency-division multiplexing (OFDM) waveforms to detect the position of a ground target under half-duplex operation. The area of interest, where the target is located, is sectioned into a grid of cells, where the radar cross-section (RCS) of every cell is jointly estimated by the UAVs, and a central node acts as a fusion center by receiving all the estimations and performing information-level fusion. For local estimation at each UAV, the periodogram approach is utilised, and a digital receive beamformer is assumed. The fused RCS estimates of the grid are used to estimate the cell containing the target. Monte Carlo simulations are performed to obtain the detection probability of the proposed framework, and our results show that the proposed framework attains improved accuracy for the detection of a target than other OFDM bi-static radar approaches proposed in the literature. 

\end{abstract}


\section{Introduction}\label{sec:Introduction}

Toward the sixth generation of wireless networks (6G), a number of exciting applications will benefit from sensing services provided by future perceptive networks, where sensing capabilities are integrated in the communication network. Once the communication network infrastructure is already deployed with multiple interconnected nodes, a multi-static sensory mesh can be enabled and exploited to improve the performance of the network itself~\cite{9296833}. Therefore, the joint communications and sensing (JCAS) concept has emerged as an enabler for an efficient use of  radio resources for both communications and sensing purposes, where high frequency bands, that are expected to be available in 6G, can favor very accurate sensing based on radar-like technology~\cite{art:Wild_Nokia}. 


Relying on the coordination of the network and a distributed processing, sensing signals can be transmitted from one node, and  the reflections on the environment can be received at multiple nodes, in a coordinated manner~\cite{art:Wild_Nokia}. Thus, distributed multi-static sensing approaches can improve sensing accuracy while alleviating the need of full-duplex operation at sensing nodes. In this context, the high-gain directional beams provided by performing beamforming in multiple-input multiple-output (MIMO) and massive MIMO systems, which are essential for the operation of communication systems at higher frequencies, will be also exploited for improving sensing by considering distributed implementations~\cite{8764485,art:multistatic_Merlano}. In multi-static MIMO radar settings, the synchronization among sensing nodes is crucial, thus, this issue has motivated the study of feasibility of synchronization. For instance, a synchronization loop using in-band full duplex (IBFD) was demonstrated for a system with two MIMO satellites sensing two ground targets in~\cite{art:multistatic_Merlano}. 

Additionally, multicarrier signals  such as orthogonal frequency-division multiplexing (OFDM) waveforms have proven to provide several advantages for the use on JSAC systems, including independence from the transmitted user data, high dynamic range, possibility to estimate the relative velocity, and efficient implementation based on fast Fourier transforms~\cite{5776640}. 
%
%
%
For instance, uplink OFDM 5G New Radio (NR) waveforms have been effectively used for indoor environment mapping in \cite{art:Baquero_mmWaveAlg}. Therein, a prototype full-duplex transceiver was used to perform range-angle chart estimation and dynamic tracking via extended Kalman Filter.

Moreover, the capabilities of distributed sensing system can be further extended by relying on the advantages of flexible nodes as unmanned aerial vehicles (UAVs), which have already raised significant attention for their applicability in numerous scenarios and even in harsh environments~\cite{8877114}. Therefore, UAVs have been already considered for sensing purposes~\cite{art:Wei_UAV_Safe,art:UAVs_Guerra,Chen_JSACUAVSystem}. 
For instance, in~\cite{art:Wei_UAV_Safe}, UAVs are explored to  perform simultaneous jamming and sensing of UAVs acting as eavesdroppers by exploiting the jamming signals for sensing purposes. Therein, sensing information is used to perform optimal online resource allocation to maximise the amount of securely served users constrained by the requirements on the information leakage to the eavesdropper and the data rate to the legitimate users. Besides in~\cite{art:UAVs_Guerra}, a UAV-based distributed radar is proposed to perform distributed sensing to locate and track malicious UAVs using frequency modulated continuous wave (FMCW) waveforms. It was shown that the mobility and distributed nature of the UAV-based radar benefits the accuracy for tracking mobile nodes if compared with a fixed radar. However, it does not make complete use of its distributed nature, as each UAV performs local sensing accounting for only the sensing information of its neighbouring UAVs, and there is no consideration on communication tasks.
In the context of JSAC, in~\cite{Chen_JSACUAVSystem}, a general framework for a full-duplex JSAC UAV network is proposed, where area-based metrics are developed considering sensing and communication parameters of the system and sensing requirements. This work uses a full-duplex operation for local sensing at each UAV while considering reflections from other UAVs as interference. 


Different from previous works and considering the complexity of full-duplex systems, this work focuses on half-duplex operation and proposes a framework for performing a grid-based distributed sensing relying on the coordination of multiple UAVs to sense a ground target located on an area of interest. It is considered that MIMO UAVs employ OFDM waveforms and digital beamforming is implemented at the receiver side. A periodogram is used for the estimation of the radar cross-section (RCS) of each cell in the grid, leveraging the knowledge of the geometry of the system. The RCS estimation is performed by all of the non-transmitting UAVs, simultaneously, while one UAV is illuminating a certain sub-area of the grid. This process is performed until all UAVs have illuminated their respective sub-areas, then all UAVs inform the measured RCS per cell on the grid to a UAV acting as a fusion center (FC), which performs information-level fusion. This process allows a half-duplex operation in a distributed sensing setting.


\section{System Model}\label{sec:SysModel}
\begin{figure}[bt]
    \centering
    \includegraphics[width=0.73\linewidth]{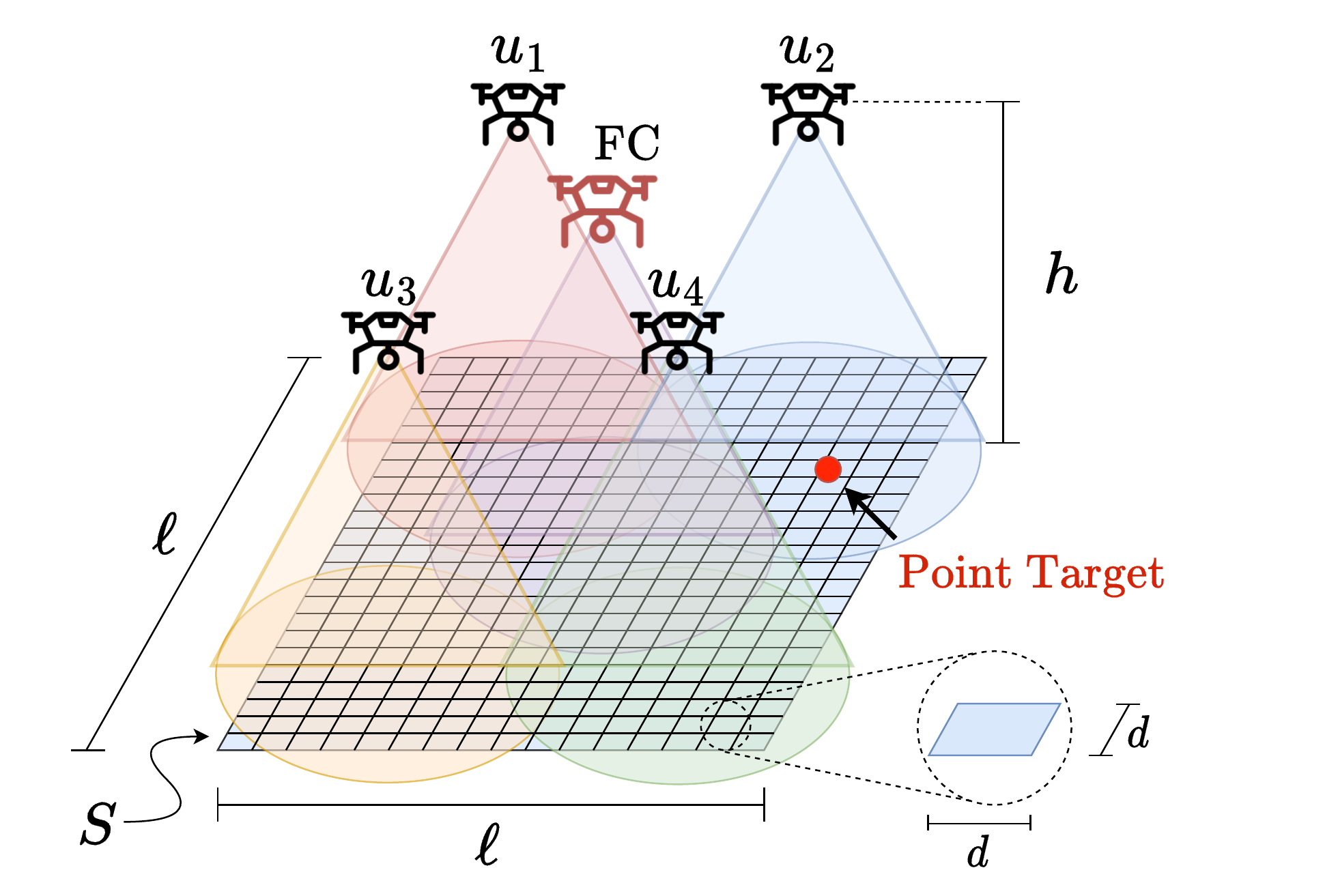}
    \caption{System model.\vspace{-1em}}
    \label{fig:sysModel}
\end{figure}
Consider the system depicted in Fig.~\ref{fig:sysModel}, where a single point-like target of RCS $\sigma_{\mathrm{T}}$ is positioned on a square area $S$ of $\ell$ meters of side length. $U$ UAVs are deployed (for simplicity and insighfulness) at a common altitude $h$ and are coordinated to perform distributed sensing to locate the ground target. Each UAV $u$ in the set of all UAVs $\mathcal{U}$, with $u\in\mathcal{U}$, is positioned at coordinates $\mathbf{r}_u = (x_u,y_u,h)$, with $|\mathcal{U}|=U$. Also, the RCS of a ground cell is denoted by $\sigma_{G}$.

Similar to~\cite{Chen_JSACUAVSystem}, it is assumed that each UAV has two arrays of antennas namely a square uniform planar array (UPA)  (mounted facing downward) for sensing  and a uniform linear array (ULA) (mounted horizontally) to communicate with the FC for information fusion and coordination tasks. The square UPA consists of $n\times n$ isotropic antenna elements spaced $\lambda/2$ from each-other, where $\lambda=f_0/c_0$ is the wavelength of the signal, $f_0$ is the frequency of the signal, and $c_0$ is the speed of light.

\if\mycmd1

To perform sensing, the UAV $u \in \mathcal{U}$ estimates the RCS of a certain point on the ground, denoted as $p$,  located at the coordinates $\mathbf{r}_p=(x_p,y_p,0)$. For this purpose, $u$ utilizes a digital receive beamformer $\mathbf{w}_p\in\mathbb{C}^{n\times 1}$. The reflection from point $p$ arriving at UAV $u$ has an angle-of-arrival (AoA) of $\varphi_{p,u} = (\theta_{p,u},\phi_{p,u})$, where $\theta_{p,u}$ corresponds to the elevation angle and $\phi_{p,u}$ to the azimuth. The corresponding beam-steering vector $\mathbf{g}(\varphi_{p,u})$ has its elements $g_{ij}(\varphi_{p,u})$ for all $i,j = 0,...,n-1$, where $i$ is the index corresponding to the antenna element in the $x$ axis and $j$ in the $y$ axis of the UPA defined as~\cite{book:BalanisAntennas}
\begin{align}\nonumber
    g_{ij}(\varphi_{p,u}) = &e^{-j\pi (i-1)\sin(\theta_{p,u})\sin(\phi_{p,u})} \times \\
    &e^{-j\pi (j-1)\sin(\theta_{p,u})\cos(\phi_{p,u})}.
\end{align}
The steering matrix $\mathbf{G}_u \in \mathbb{C}^{n^2\times H}$ contains the steering vectors corresponding to the $H$ reflections captured at UAV $u$ as
\begin{equation}
    \mathbf{G}_u = [\mathbf{g}(\varphi_{1,u}),..., \mathbf{g}(\varphi_{H,u})]_{n^2\times H},
\end{equation}
where $n^2$ is the total number of antennas at UAV $u$. 

After applying the receive beamformer $\mathbf{w}_{p}$ at reception, the beam pattern from the reflections captured at $u$ is given by
\begin{align}
    \pmb{\chi} =  \mathbf{G}_u^T\mathbf{w}_{p} = [\chi(\varphi_{1,u}),...,\chi(\varphi_{H,u})]^T,
\end{align}
where $\chi(\varphi_{p,u})$ is the gain of the reflection coming from $p$ and $\pmb{\chi}$ is the beam pattern vector of size $H\times 1$ at every AoA  by applying beamformer $\mathbf{w}_{p}$ at UAV $u$.

\fi

\section{Distributed Sensing Protocol}\label{sec:Protocol}

\begin{figure}[bt]
    \centering
    \includegraphics[width=0.95\linewidth]{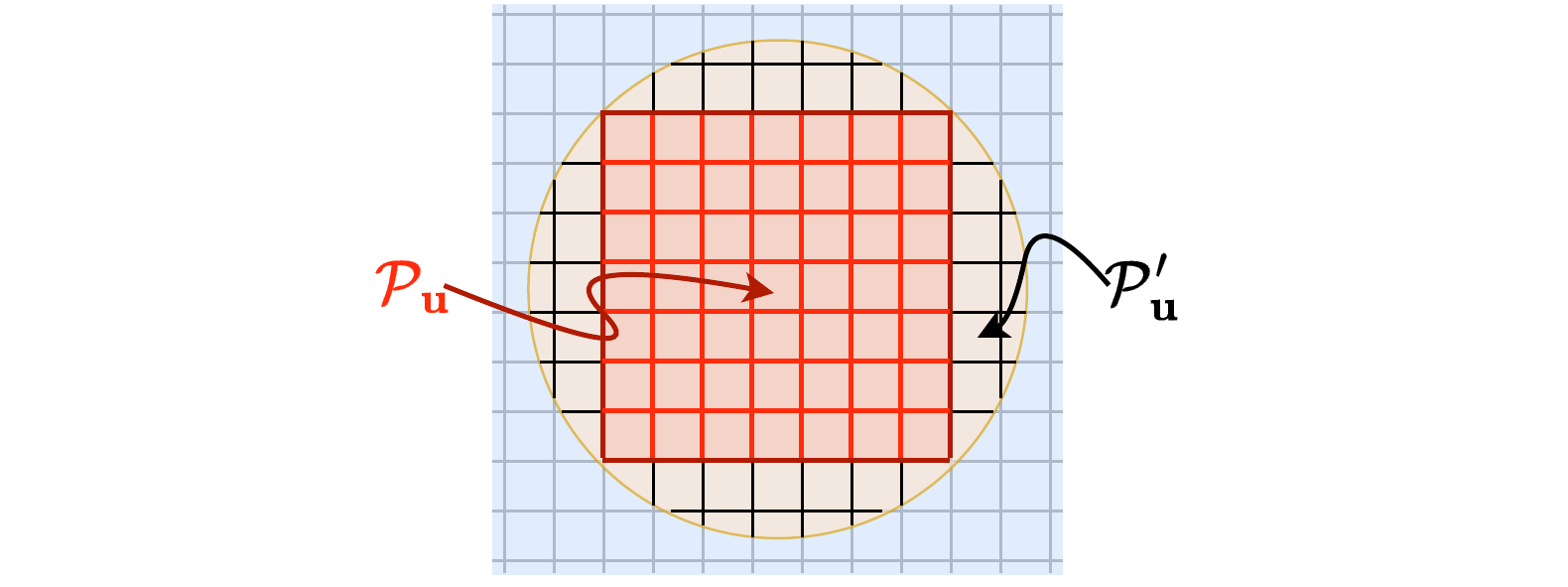}
    \caption{Illumination grid.}
    \label{fig:gridCells}
    \vspace{-1em}
\end{figure}

For the sensing process, it is considered that the total area $S$ is sectioned into a grid composed of $L\times L$ square cells with dimensions $ d \times d $ with $d = \ell/L$. Each cell is characterised by its middle point $p$ of position $\mathbf{r}_p =(x_p,y_p,0)$ such that $p\in\mathcal{P}$, where $\mathcal{P}$ is the set of all cells. For notational simplicity we will refer a certain cell by its middle point $p$. The point $p^*$ represents the target, which is located in the position $\mathbf{r}_{p^*} =(x_{p^*},y_{p^*},0)$.
It is also considered that, at a certain time, a UAV $u\in\mathcal{U}$ illuminates straight down with its UPA working as a phased array, 
thus the half power beam width (HPBW) projection forms a circle on the ground. In this sense, it is assumed that the cells that are completely inside the largest inscribed square of the HPBW projection are the intended ones to be sensed by the reflections produced from the illumination of UAV $u$, and are characterised as the cell set $\mathcal{P}_{u}$, while the set of non-intended illuminated cells $\mathcal{P}_{u}'$ contains the cells that are not inside the largest inscribed squared, which are treated as clutter, as illustrated in Fig.~\ref{fig:gridCells}.
In total, the set of illuminated cells 
is given as $\mathcal{Q}_{u} = \mathcal{P}_{u}\cup\mathcal{P}_{u}'$.

The distributed sensing framework is summarized as follows
\begin{itemize}[label={},leftmargin=*]
    \item \textbf{Step 1:} The UAVs coordinate and assume their positions to cover the whole area of interest $S$, such that every cell in $\mathcal{P}$ is contained in a single $\mathcal{P}_u$, $u\in\mathcal{U}$.
    \item \textbf{Step 2:} UAV $u\in\mathcal{U}$ illuminates the ground directly below acting as a phased array, illuminating the elements of $\mathcal{Q}_u$, and potentially, the target $p^*$. 
    \item \textbf{Step 3:} Every other UAV  $u'\in\mathcal{U}\setminus\{u\}$ processes the incoming reflections by choosing a cell $p\in\mathcal{P}_u$ and for that cell
    \begin{itemize}
        \item computes and applies a digital receive beamformer as described in Section~\ref{sec:BF}, and
        \item computes the periodogram corresponding to $p$, and estimates its RCS as described in Section~\ref{sec:Periodogram}.
    \end{itemize}
    
    \item \textbf{Step 4:} Repeat Step 3 for all cells $p\in\mathcal{P}_u$.
    \item \textbf{Step 5:} Repeat Steps 2-4 for all UAVs $u\in\mathcal{U}$. After this, each UAV $u$ has an estimated RCS map of the grid, $\mathbf{\Hat{\Gamma}}_u$, which is a matrix of RCS estimates of all cells in $\mathcal{P} \setminus \mathcal{P}_u$. This occurs because the UAV $u$ does not estimate the RCS of the cells in $\mathcal{P}_u$, thus avoiding the need for a full-duplex system at the UAVs.
    \item \textbf{Step 6:} All UAVs $u\in\mathcal{U}$ send their RCS estimation maps $\mathbf{\Hat{\Gamma}}_u$ to the FC for information-level fusion.
    \item \textbf{Step 7:} The FC fuses the estimates together into the fused RCS map $\mathbf{\Hat{\Gamma}}$, and, by assuming a non-reflective ground such that the RCS of the ground is smaller than that of the target ($\sigma_{\mathrm{G}} < \sigma_{\mathrm{T}}$), the target is estimated to be located in the cell of highest estimated RCS, i.e. in $\argmax \mathbf{\Hat{\Gamma}}$, as described in Section~\ref{sec:Fuse}.
\end{itemize}



\section{Beamformer Design}\label{sec:BF}
The receive beamformer is designed to have the main lobe of the resulting beam pattern steered towards the intended cell $p$ in order to estimate its RCS. To this end, two different approaches are considered for the design of the receive beamformer, namely least-squares (LS) heuristic formulation and the minimum variance beam pattern based on Capon method. These approaches are described in the following.

\subsection{Least-Squares heuristic approach}

For this approach, the beamformer is obtained by solving the following constrained LS optimisation problem \cite{art:Shi_ILS,art:Zhang_DBR} 
\begin{subequations}
\begin{alignat}{3}
    \mathrm{\textbf{P1:}}\;\;\;\;&\mathrm{minimise}&\qquad&|| \mathbf{G}^T\mathbf{w}_{p} - \mathbf{v} ||_2^2\\
    &\mathrm{subject~to}&\qquad&|| \mathbf{w}_{p} ||_2^2 = 1,
\end{alignat}
\end{subequations}
where $\mathbf{v}$ is the desired response vector over the $H'$ AoAs in the beam-steering matrix $\mathbf{G}\in\mathbb{C}^{n\times H'}$. 
In this approach, a heuristic is employed, in which the AoAs in $\mathbf{A}$ are chosen such that they span evenly on the elevation and azimuth ranges, centred around the intended AoA $\varphi_{p,u}$. The AoAs are taken as a mesh of $n$ elevation angles and $4n$ azimuth angles respectively given by
\begin{alignat}{3}
    \theta_i =&\!\mod{\left( \theta_{p,u} + \frac{i\pi}{2(n-1)}, \frac{\pi}{2} \right)}, \; & i=0,...,n-1\\
    \phi_j =&\!\mod{\left( \phi_{p,u} + \frac{j2\pi}{4n-1}, 2\pi \right)}, \; & j=0,...,4n-1,
\end{alignat}
such that $H' = 4n^2$.

The solution of \textbf{P1} is well known to be $\mathbf{w}_{p} = (\mathbf{A}^T)^\dagger \mathbf{v}$ \cite{art:Zhang_DBR} where $(\cdot)^\dagger$ is the pseudo-inverse, but, since $\mathbf{A}^T$ is a matrix with more rows than columns, it can be efficiently solved by applying Cholesky factorization. Therefore, the iterative LS algorithm proposed in \cite{art:Shi_ILS} can be employed to solve \textbf{P1}.

\subsection{Capon method}

The Capon method provides minimum-variance distortionless
response beamforming and can be formulated as a quadratic program (QP) convex optimisation problem \cite{art:Stoica_Capon}
\begin{subequations}
\begin{alignat}{3}
\mathrm{\textbf{P2:}}\;\;\;\;&\mathrm{minimise}&\qquad&\mathbf{w}_{p}^H\mathbf{R}\mathbf{w}_{p}\\
    &\mathrm{subject~to}&\qquad& \mathbf{w}_{p}^H\mathbf{g}(\varphi_{p,u'})  = 1,
\end{alignat}
\end{subequations}
where $\mathbf{R}\in\mathbb{C}^{n\times n}$ is the covariance matrix of the received signal over the desired direction, which can be defined as $\mathbf{R} = \mathbf{g}(\varphi_{p,u'})\mathbf{g}(\varphi_{p,u'})^H + \alpha \mathbf{I}$~\cite{art:Shi_ILS} , 
where $\mathbf{I}\in\mathbb{R}^{n\times n}$ is the identity matrix and $\alpha$ is a small real number. The solution for \textbf{P2} is obtained as in~\cite{art:Stoica_Capon}, and given by
\begin{equation}
    \mathbf{w_{p}} = \frac{\mathbf{R}^{-1}\mathbf{g}(\varphi_{p,u'})}{\mathbf{g}(\varphi_{p,u'})^H\mathbf{R}^{-1}\mathbf{g}(\varphi_{p,u'})}.
\end{equation}

\section{Periodogram}\label{sec:Periodogram}
\if\mycmdPeriodogram0

\fi
\if\mycmdPeriodogram1
For performing sensing, the UAVs illuminating ground transmit frames consisting of $N$ OFDM symbols, each consisting of $M$ orthogonal subcarriers. The transmitted OFDM frame can be expressed as a matrix denoted by $\mathbf{F_{TX}}=[c^{TX}_{k,l}]\in\mathcal{A}^{N\times M}$ with $k=0,...,N-1$, $l=0,...,M-1$ and $\mathcal{A}$ is the modulated symbol alphabet.  At the side of sensing UAVs, the received frame matrix is denoted by $\mathbf{F_{RX}}=[c^{RX}_{k,l}]$ and is composed by the received baseband symbols corresponding to all reflections from $\mathcal{Q}_{\mathrm{u}}$ at UAV $u'$. The elements of the received frame matrix have the form~\cite{art:Baquero_OFDM,art:OFDM_Samir}
\begin{align}\label{eq:bPointTarg_Mult} \nonumber
    &c^{RX}_{k,l} =   ~~b_{p}c^{TX}_{k,l}\chi(\varphi_{p,u'}) e^{j2\pi f_{D,p}T_o l}e^{-j2\pi \tau_{p} \Delta f k}e^{-j\zeta_{p}}\\ \nonumber
            &+\sum_{p'\in\mathcal{Q}_{\mathrm{u}}\setminus\{p\}} b_{p'}c^{TX}_{k,l}\chi(\varphi_{p',u'}) e^{j2\pi f_{D,p'}T_o l}e^{-j2\pi \tau_{p'} \Delta f k}e^{-j\zeta_{p'}}\\ 
            &+ \delta_{u}b_{p^*}c^{TX}_{k,l}\chi(\varphi_{p^*,u'}) e^{j2\pi f_{D,p^*}T_o l}e^{-j2\pi \tau_{p^*} \Delta f k}e^{-j\zeta_{p^*}}+ z_{k,l}, 
\end{align}
where $f_{D,p}$ is the Doppler experienced by the reflection from $p$ (assumed constant through the frame), $T_o$ is the OFDM symbol duration (including the cyclic prefix time $T_{CP}$), $\tau_p$ is the delay of the reflection from $p$, $\Delta f$ is the inter-carrier spacing, $\zeta_p$ is a random phase shift of the reflection from $p$, $z_{k,l}$ is additive white Gaussian noise (AWGN) of spectral density $N_0$, and $b_p$ is a term embedding the propagation of the wave and the reflecting characteristics of the reflector in $p$. In this expression, the first term corresponds to the intended cell $p$; the second term corresponds to the interference from the other cells, $p'$, in $\mathcal{Q}_u$; and the third term corresponds to the target reflection, with $\delta_{u}=1$ if the target has been illuminated by UAV $u$, and $\delta_{u}=0$ otherwise. Considering a point-source model, $b_p$ is the amplitude attenuation of the signal, given by~\cite{art:OFDM_Samir}
\begin{equation}\label{eq:b_val}
    b_p = \sqrt{\frac{P_T G_T \sigma_p \lambda^2}{(4\pi)^3 d_{p,1}^2d_{p,2}^2}},
\end{equation}
where $\sigma_p\in\{\sigma_{\mathrm{G}},\sigma_{\mathrm{T}}\}$, $P_T$ is the transmit power, $G_T$ is the transmit antenna gain, $d_{p,1}$ is the distance from $u$ to $p$ and $d_{p,2}$ is the distance from $p$ to $u'$.

The received complex symbols $c^{RX}_{k,l}$ contain the transmitted symbols $c^{TX}_{k,l}$, thus, are data-dependent. In order to process $\mathbf{F_{RX}}$, this data-dependency is removed by performing element-wise division, $\mathbf{F}$$=$$\mathbf{F_{RX}}\oslash \mathbf{F_{TX}}$,  to obtain the processed samples consisting of elements $c_{k,l}$$ = $$c^{RX}_{k,l}/c^{TX}_{k,l}$. 

To estimate the delay and Doppler from $\mathbf{F}$, a common approach for OFDM signals is to use the periodogram, which provides the maximum likelihood (ML) estimator~\cite{art:Baquero_OFDM}. The periodogram takes the fast Fourier transform (FFT) of $\mathbf{F}$ over OFDM symbols, followed by the inverse FFT (IFFT) over subcarriers at a given delay-Doppler pair $(n,m)$ as~\cite{art:Baquero_OFDM}
\begin{align}\label{eq:periodogram}
    \nonumber P&_{\mathbf{F}}(n,m) = \\
    &\frac{1}{NM} \left| \sum_{k=0}^{N'-1}\left( \sum_{l=0}^{M'-1} c_{k,l} e^{-j2\pi l\frac{m}{M'}} \right) e^{-j2\pi k\frac{n}{N'}} \right|^2,
\end{align}
where $M'\geq M$ and $N'\geq N$ are the lengths of the FFT and IFFT operations respectively, $n=0,...,N'-1$ and $m=0,...,M'-1$\footnote{If $M' > M$ or $N' > N$ is needed in order to have more $n$ or $m$ values, padding is used by setting the added padded symbols to zero.}. 
It has been proven that the ML estimator of the delay and Doppler for a single target coincides with the maximum point in the periodogram as $(\hat{n}, \hat{m}) = \argmax_{n,m} P_{\mathbf{F}}(n,m)$~\cite{art:Baquero_OFDM}, which is maximised when
\begin{align}\label{eq:periodogramMax}
    \frac{f_D}{\Delta f} -\frac{\hat{m}}{M} = 0 \;\;\;\land\;\;\;    \frac{\tau}{T_o} -\frac{\hat{n}}{N} =0.
\end{align}

Then, from \eqref{eq:bPointTarg_Mult}, \eqref{eq:b_val} and \eqref{eq:periodogram}, $\sigma_p$ can be estimated as
\begin{equation}\label{eq:rcsestcalc}
    \Hat{\sigma}_p = \left(\frac{1}{NM}\right) \frac{P_{\mathbf{F}}(\hat{n}, \hat{m})(4\pi)^3d_{p,1}^2d_{p,2}^2}{P_TG_T\lambda^2}.
\end{equation}
Then, considering the the geometry and protocol of the system, each UAV can set $\hat{m}$, $M$, $\hat{n}$ and $N$ so that \eqref{eq:periodogramMax} is met exactly for each cell $p$ to be sensed, and the corresponding RCS estimate is obtained by computing \eqref{eq:rcsestcalc}.
\fi

\section{Information-Level Fusion}\label{sec:Fuse}

After all UAVs $u\in\mathcal{U}$ have sent their local RCS estimates for each cell on the grid $\mathbf{\Hat{\Gamma}}_{u}$ to the FC, it performs information-level fusion of the local estimates to obtain a global estimate $\mathbf{\Hat{\Gamma}}$. Then, the following hypothesis test is performed for all the elements of the grid, $p\in\mathcal{P}$
\begin{subequations}\label{eq:hypothesis}
\begin{alignat}{3}
\mathcal{H}_0:\;\;\;\;&|| \mathbf{r}_{p^*} - \mathbf{r}_{p} ||_\infty > \frac{d}{2}  \\
\mathcal{H}_1:\;\;\;\;&|| \mathbf{r}_{p^*} - \mathbf{r}_{p} ||_\infty \leq \frac{d}{2},
\end{alignat}
\end{subequations}
where $||\cdot||_\infty$ is the $L^\infty$ norm. Hypothesis $\mathcal{H}_1$ states the cases when the target $p^* $ is considered to be located in the corresponding cell $p$, and is considered to be met at the cell $p$ that has the maximum value estimate $\Hat{\sigma}=\max \mathbf{\Hat{\Gamma}}$. On the other hand, $\mathcal{H}_0$ states the cases when the target is not located at $p$, and considered to be met at every other cell $p$ that has an estimate $\Hat{\sigma}$ such that $\Hat{\sigma} < \max \mathbf{\Hat{\Gamma}}$. 

The information-level fusion will be carried out using two techniques, namely averaging and pre-normalising the local estimates before averaging.

\textbf{Average: }
The FC averages the values of the cells over the local maps from all UAVs in $\mathcal{U}$ such that $\mathbf{\Hat{\Gamma}} = \frac{1}{U}\sum_{u\in\mathcal{U}} \mathbf{\Hat{\Gamma}}_{u}$.

\textbf{Pre-normalised average: } An average of the pre-normalised local maps is obtained, in which each local map $\mathbf{\Hat{\Gamma}}_{u}$ is normalised between 0 and 1 as
\begin{equation}
    \Bar{\sigma} = \frac{\Hat{\sigma} - \min{(\mathbf{\Hat{\Gamma}}_{u})}}{\max{(\mathbf{\Hat{\Gamma}}_{u})} - \min{(\mathbf{\Hat{\Gamma}}_{u})}}\enskip,\enskip \forall \Hat{\sigma}\in\mathbf{\Hat{\Gamma}}_{u} \enskip,\enskip \forall u \in\mathcal{U}.
\end{equation}

The resulting normalised local maps $\Bar{\Gamma}_{u}$ are then averaged as in the previous approach.

\section{Numerical Results}\label{sec:Results}

In this section, the performance of the proposed sensing protocol will be evaluated in terms of the probability of detection, where detection is considering to occur whenever $\mathcal{H}_1$ is achieved for the cell that contains the target. For this purpose, Monte Carlo simulations were performed, where the target is randomly located at each iteration, and the simulation parameters are presented in Table~\ref{tab:commonPar}, unless stated otherwise. The value for $\sigma_\mathrm{T}$ is assumed as 10 dBsm\footnote{$\sigma$[dBsm] = 10$\log_{10}(\sigma$[$m^2$]$/1 m^2)$}, which is a reasonable value for large vehicles~\cite{art:RCSvals01}, and -30 dBsm for the ground, which is reasonable for grass or dirt~\cite{art:RCSvalsGND}. The OFDM parameters are taken from \cite{Chen_JSACUAVSystem}. The UAVs are set uniformly distributed across $S$ in a way to cover the whole grid, and each one illuminates $L/\sqrt{U}\times L/\sqrt{U}$ cells and avoiding intersections between the $\mathcal{P}_{u}$ cells, unless stated otherwise.
\begin{table}[h!]\vspace{-1em}\centering
\caption{Common simulation parameters}\label{tab:commonPar}
  \begin{tabular}{|c|c||c|c|}
    \hline
    \textbf{Parameter}  & \textbf{Value}    &   \textbf{Parameter}  & \textbf{Value}       \\ \hline
    $P_T$                   & 1 [W]         &   $M$                   & 64        \\ \hline
    $G_T$                   & 1             &   $n$                     & 8         \\ \hline
    $\ell$                     & 100 [m]       &   $f_0$                   & 24 [GHz]  \\ \hline
    $U$                     & 16            &   $BW$                    & 200 [MHz] \\ \hline
    $N_0$                   & -174 [dBm/Hz] &   $T_{CP}$                & 2.3 [$\mu$s] \\ \hline
    $\sigma_{\mathrm{G}}$   & -30 [dBsm] &   $L$                     & 20            \\ \hline
    $\sigma_{\mathrm{T}}$   & 10 [dBsm]    &   $f_D$                   & 0 [Hz]        \\ \hline
    $N$                   & 16            &                      &           \\ \hline
  \end{tabular}
   \vspace{-1em}
\end{table}

\begin{figure}
    \centering
    \includegraphics[width=0.73\linewidth]{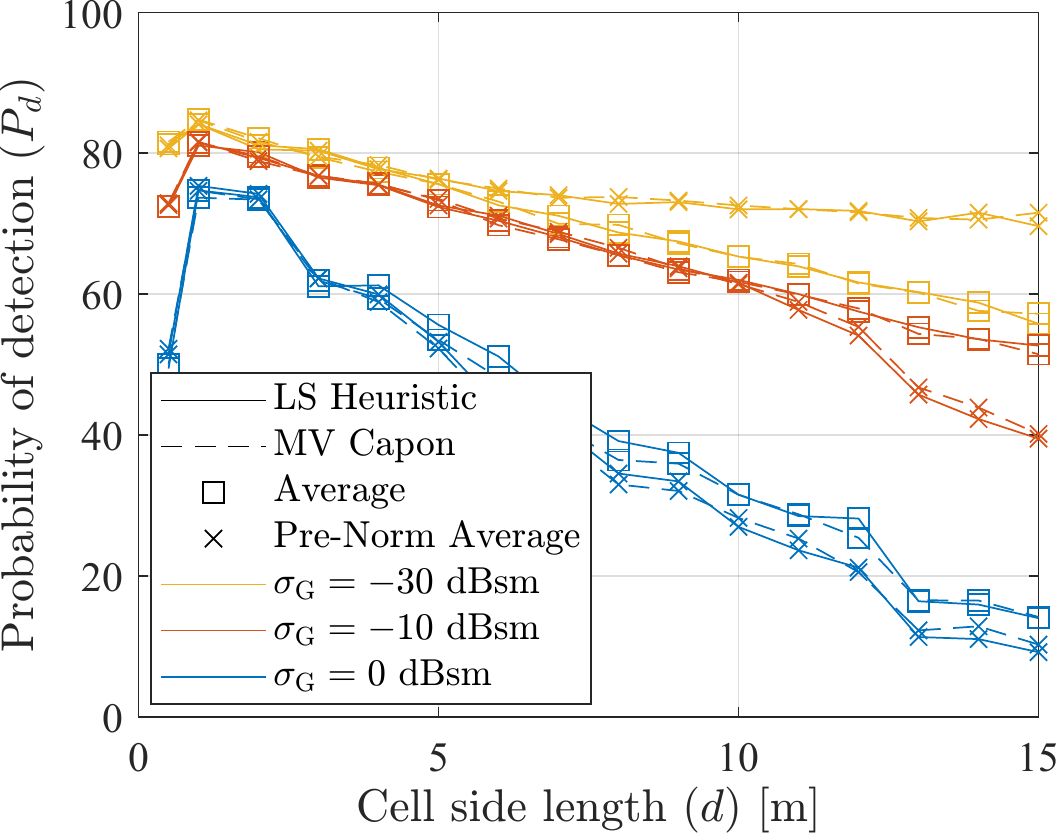}
    \caption{Detection probability of the target for different cell length $d$ for different beamforming techniques, fusion techniques and $\sigma_{\mathrm{G}}$ values. The number of UAVs and number of cells illuminated per UAV is kept constant, so larger $d$ values imply total area and higher UAV altitude.\vspace{-1em}}
    \label{fig:resultNew_Pd_d_BF}
\end{figure}

\begin{figure}
    \centering
    \includegraphics[width=0.73\linewidth]{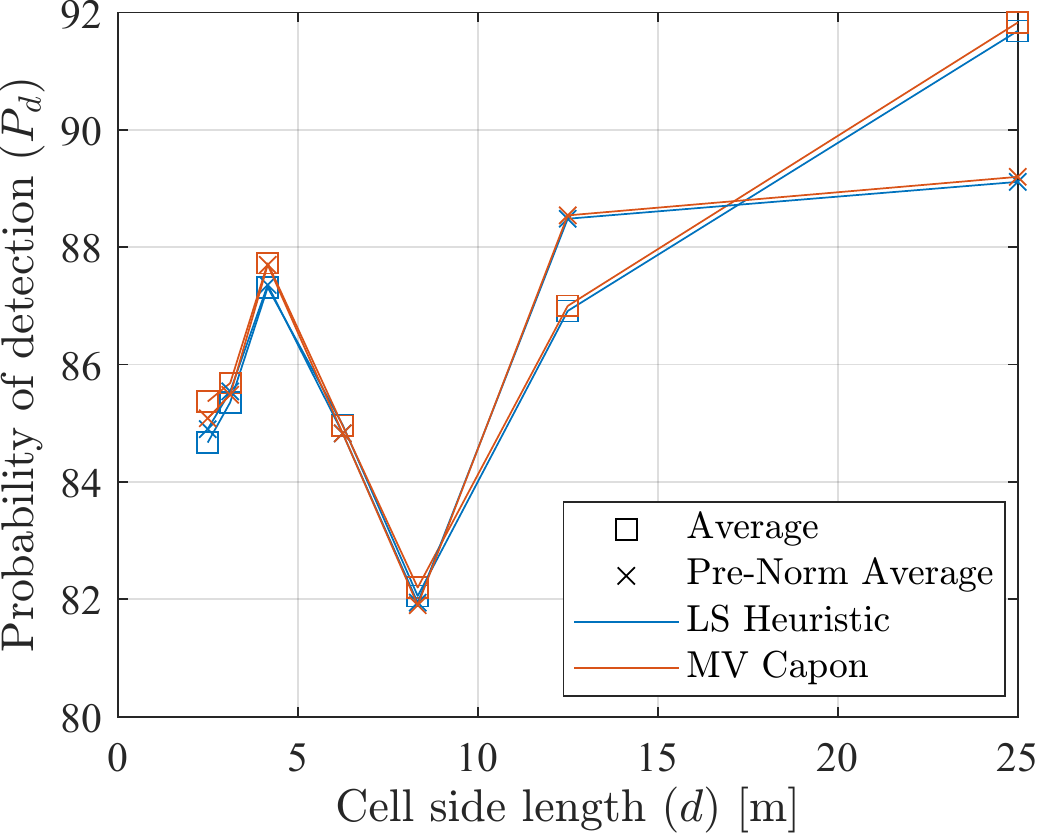}
    \caption{Detection probability of the target for different cell length $d$ for different beamforming and fusion techniques. Total area and UAV altitude is kept constant, so larger $d$ values imply less cells illuminated per UAV.\vspace{-2em}}
    \label{fig:resultNew_Pd_d_VAR}
\end{figure}


In Fig.~\ref{fig:resultNew_Pd_d_BF} the detection probability is shown as a function of the cell side length $d$, for different $\sigma_{\mathrm{G}}$ values. The number of intended cells per UAV is maintained constant, thus the size of the cells determine the total size of the area $S$ and the altitude of the UAVs $h$, which increases as $d$ increases to accommodate the same number of larger cells in its HPBW. Note that, $d$ increases as $h$ increases, and the $b_p$ value from \eqref{eq:b_val} is closer to the noise level, then the probability of detection decreases. 
There exists a maximum point around $d=2$m, where the best probability of detection is achieved.
As expected, as $\sigma_{\mathrm{G}}$ increases, the difference between the RCS of the ground and the target decreases, so that the probability of detection also decreases. By comparing beamforming techniques, both show a similar behaviour. However, when comparing fusion techniques, pre-normalising the local estimates performs better only for larger $d$ and $\sigma_{\mathrm{G}}$ values.

Conversely, in Fig.~\ref{fig:resultNew_Pd_d_VAR} the total size of the area $S$ and the altitude of the UAVs $h$ is kept constant, while varying $d$. This is accomplished by adjusting the number of cells in the grid $L$. 
In this case, note that higher $d$ values lead to better probability of detection, as there is a higher area per cell. However, a local optimum can be appreciated around $d=4$m, which shows the presence of a local optimum that offers more precise detection.

\begin{figure}
    \centering
    \includegraphics[width=0.73\linewidth]{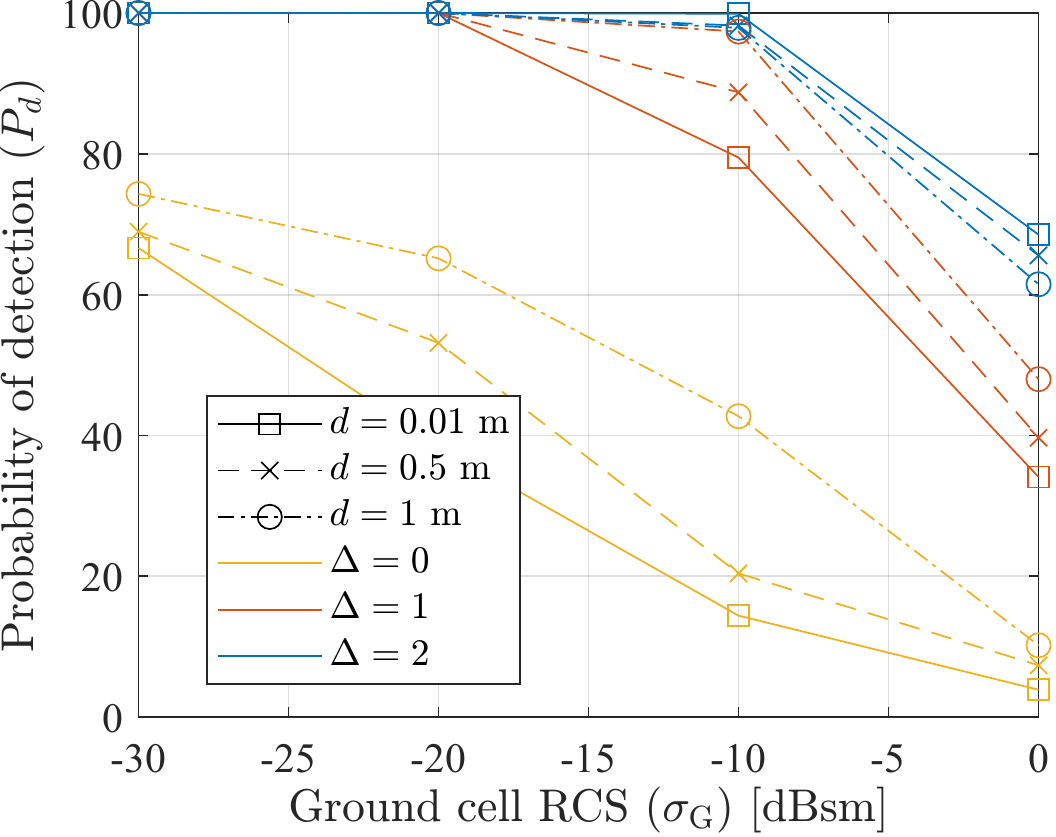}
    \caption{Detection probability of the target at a $\Delta$ cells distance for different $\sigma_{\mathrm{G}}$ values and different cell size $d$.\vspace{-1em}}
    \label{fig:resultNew_Pd_Dd}
\end{figure}

Furthermore, in Fig.~\ref{fig:resultNew_Pd_Dd} the detection probability is plotted for different values of $\sigma_{\mathrm{G}}$, different values of a modified threshold $d(\frac{1}{2} + \Delta$) in the hypothesis test \eqref{eq:hypothesis}, and different values of $d$. The curves show the probability of detection of the target at $\Delta$ cells away from the cell with the maximum value in $\Bar{\Gamma}$. Note that values of $\sigma_{\mathrm{G}}\leq -10$dBsm present probability of detection close to 100\% for $\Delta \geq 1$, which is within a distance of one cell. This suggests a probability of detection above 99.89\%  with high accuracy within $5$cm ($d=0.01$m, $\Delta=2$, $\sigma_{\mathrm{G}}\leq -10$dBsm), which is more accurate than other state-of-the-art works utilising MIMO OFDM bistatic-radar such as in~\cite{art:Lyu_MIMOOFDMEurasip}, where they achieve an accuracy of $3$m which uses passive radar in a multi-user MIMO OFDM setting. The results show that for small $\sigma_{\mathrm{G}}$ values, most misdetections occur in an adjacent cell.


\begin{figure}
    \centering
    \includegraphics[width=0.73\linewidth]{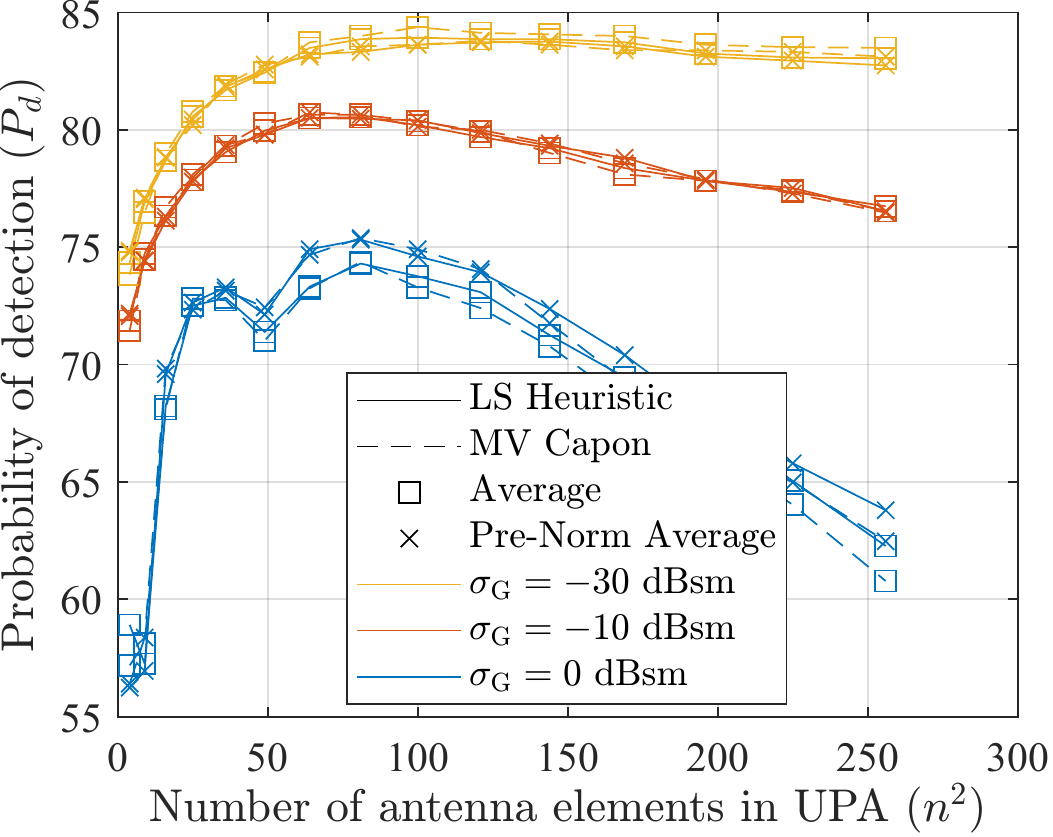}
    \caption{Detection probability of the target for different total number of antennas for the UAV UPA $n^2$ for different beamforming techniques, fusion techniques and $\sigma_{\mathrm{G}}$ values.\vspace{-2em}}
    \label{fig:resultNew_Pd_n_BF}
\end{figure}




Fig.~\ref{fig:resultNew_Pd_n_BF} illustrates the detection probability as a function of the number of antennas in the UPAs of the UAVs, for different $\sigma_{\mathrm{G}}$ values. Therein, the number of UAVs and the number of illuminated cells per UAV is maintained constant, so that narrower beams imply that the UAVs increase their altitudes to accommodate the same intended cells. It is worth noting that the increase of the number of antennas derives into a narrower main beam, and as the beam becomes narrower (higher $n$ values), it is observed an improvement on the probability of detection due to the increased directionality and precision towards the intended sensed cells. However when the beam becomes too narrow, asmall beam misalignment have a bigger impact on the detection of the target, and the increases in the UAV altitudes causes a stronger pathloss, making the received signal closer to the noise level, thus the probability of detection decreases. 
For larger  $\sigma_{\mathrm{G}}$ values, the probability of detection decreases even further, as expected. It is also noticed that both fusion techniques show a similar detection probability results, similar to the case with both beamforming techniques. However, the Capon method shows a slightly better performance for a high number of antennas and a small $\sigma_{\mathrm{G}}$ value. Moreover, for smaller $\sigma_{\mathrm{G}}$ values, the fusion by averaging slightly outperforms the pre-normalised averaging approach, while for higher $\sigma_{\mathrm{G}}$ values the opposite is true.

\begin{figure}
    \centering
    \includegraphics[width=0.73\linewidth]{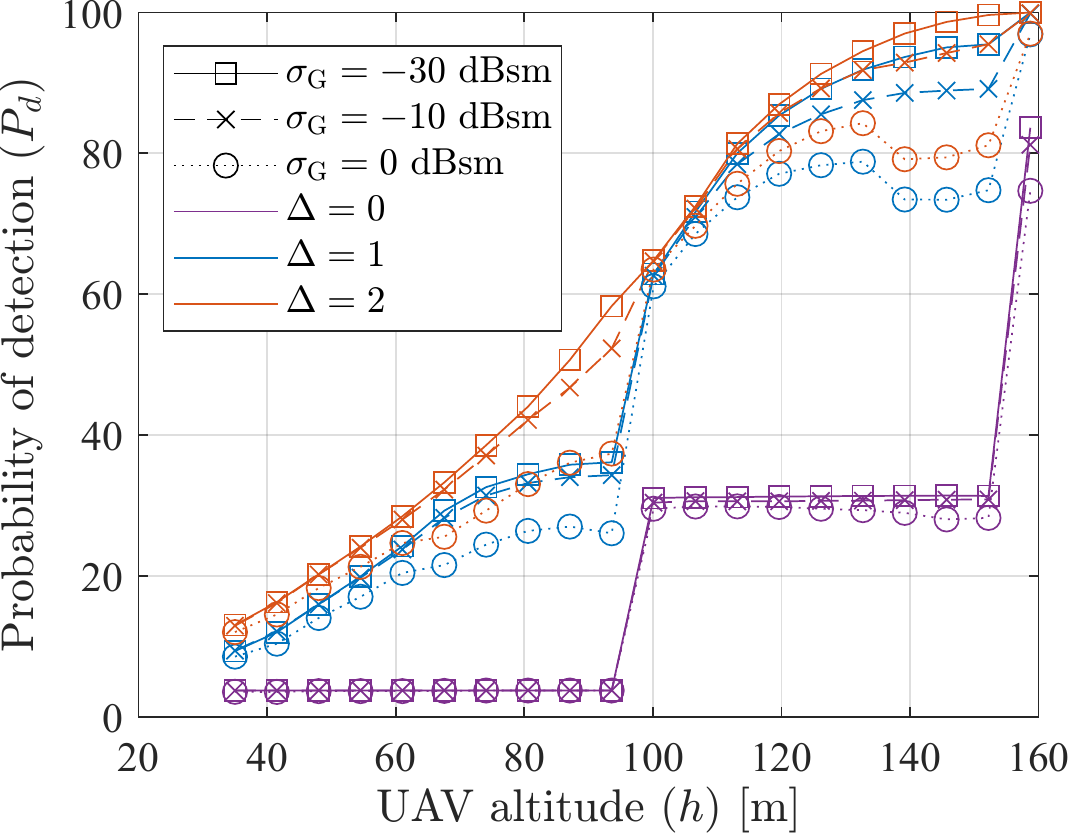}
    \caption{Detection probability of the target at a $\Delta$ cells distance for different $\sigma_{\mathrm{G}}$ values for different UAV altitude $h$ values.\vspace{-2em}}
    \label{fig:result_Pd_zUAV}
\end{figure}

Fig.~\ref{fig:result_Pd_zUAV} illustrates the detection probability as a function of the common UAV altitude $h$ for varying $\sigma_{\mathrm{G}}$ and $\Delta$ values. The UAVs are positioned in a similar configuration to the previous figure, thus, less cells are covered by the main beam of the transmitting UAVs at smaller altitude, thus resulting in cells not being illuminated by any UAV. The maximum altitude is considered to be the one where all cells are illuminated once (no overlapping). As the $h$ value increases, each $\mathcal{P}_{u}$ set goes from allocating $1\times 1$ cell, to $3\times 3$ cells and finally to $5\times 5$ cells, such that all cells are illuminated once. This behaviour can be seen in the $\Delta=0$ curve, where a sudden increasing in the probability of detection is observed at altitudes where more cells are allocated in $\mathcal{P}_{u}$, whereas this tendency is also observed for higher $\sigma_{\mathrm{G}}$ values, with worse performance. For higher $\Delta$ values, the probability of detection is higher and increases smoothly with $h$ as higher $\Delta$ implies that the detection can be considered as successful on non-illuminated cells that are adjacent to illuminated cells. This is particularly observed for $\Delta=2$, where every cell in the grid is considered for detection.


\vspace{-1em}
\section{Conclusions}\label{sec:Conclusions}
\vspace{-0.2em}
In this paper, a half-duplex distributed sensing framework for UAV-assisted networks was proposed 
in which the area of interest is sectioned into a grid, and the RCS of each cell is estimated by employing receive digital beamforming and a periodogram-based approach, and later sent to a FC for information-level fusion. Results show that the detection probability of the system increases for ground cells of smaller RCS values and that higher accuracy can be achieved within a one-cell distance. Increasing the number of antennas in the UAVs improves the detection probability of the target, however the increase of the altitude of the UAVs can deteriorate it. Moreover, it was found that detection probability is higher for larger cell size $d$ if the UAV altitude is kept constant, however there is a small $d$ value local maximum. Future works can consider the effect of the Doppler and position control of the UAVs to increase the sensing performance of the framework.

\section*{Acknowledgement}
This research has been supported by the Academy of Finland, 6G Flagship program under Grant 346208 and project FAITH under Grant 334280.
\bibliographystyle{unsrt}  
\bibliography{references}

\end{document}